**The glass transition of hyperquenched glassy water.**

V. Velikov, S. Borick and C. A. Angell
Dept of Chemistry and Biochemistry,
Arizona State University, Tempe, AZ 85287-1604

**Abstract.**

Water, water, dew divine,
An everchanging tale is thine.
Our every theory dost deny,
O, pray, where DOTH thy Tg lie!

**We review the search for the glass transition in water in its various amorphous forms, and highlight the paradoxes that the search has produced. Focussing on the glassy form of water obtained by hyperquenching, we examine its reported calorimetric properties in the light of new findings for bulk glassformers vitrified by hyperquenching. Like the hyperquenched bulk glassformers, hyperquenched water exhibits a large relaxation exotherm on reheating, as the high enthalpy quenched form relaxes to a lower enthalpy state. When the scanning temperature is scaled by the glass transition temperature, the bulk glassformers show a common pattern in this relaxation exotherm. However, when the relaxation exotherm for hyperquenched water is scaled onto the plot for the bulk glassformers using the commonly accepted glass transition temperature for water, 136K, its behavior appears completely different. If we require that the behavior of water be comparable to the other cases, except for crystallizing before the glass transition (like the majority of hyperquenched metallic glasses), then the glass transition for water must be reset to a value of 165-170K.**

As a source of scientific dissension, liquid water is as unrivalled as it is abundant. In terms of abundance in the universe, however, it is not the liquid but the glassy state of water that is dominant [1]. This substance, which comprises the body of the typical comet, is commonly formed by the slow accumulation of water molecules from the vapor state on to a cold substrate [1]. The substrates in the vastness of space are dust particles, which later agglomerate into larger bodies, such as comets. In the laboratory the cold substrate is more likely to be a differential thermal analysis DTA pan [2] a differential scanning calorimeter DSC pan [3] or an infrared IR- transparent window [4] but the substance formed has the same spectroscopic properties as the extraterrestial material [1].

This vapor deposited-material has been called amorphous solid water ASW[4], and it has been the subject of study since 1935 when Burton and Oliver [5] first showed its x-ray amorphous character.
Since it has the appearance of a glass when carefully prepared, it has been natural to inquire after the value, for water, of the usual characterizing property of the glassy state, which is its glass transition temperature. It is the fluctuating and uncertain answer to this question, over 50 years of controversy, that is the focus of the present contribution.

The subject of the glass transition temperature of the amorphous solid form of water has a checkered history. Pryde and Jones [6] were the first to deliberately seek to establish its value, and they were disappointed by their failure to find any way of assigning it. The usual thermal signature, which is the relatively sudden onset of an increased heat capacity associated with the onset of diffusion and fluidity [7], was not found within the precision of their experiment (except for one initial and irreproducible experiment).. Ghormley, who performed several characterizing experiments on the vapor deposited solid [8], also failed to find a thermal effect, though he obtained values of the heat of crystallization that are quite close to the presently accepted value of 1.29 $Jmol^{-1}K^{-1}$ [9]. Shortly thereafter, MacMillan and Los [2] prepared glassy water by direct deposit onto the pan of their vacuum-mounted differential thermal analysis apparatus, and observed a well-defined endotherm at 139K. This was apparently confirmed by the subsequent study of Sugasaki et al [10] who deposited ASW in an adiabatic calorimetry cell and repeatedly observed an endothermic effect which, unfortunately, was never completed before crystallization of the sample occurred.

Olander and Rice [11] refined the deposition process and the transmission IR spectra they recorded suggested that crystallization of the glass-like deposit never occurred below ~160K. This was confirmed in subsequent studies from Rice's group [4]. Encouraged by the increased temperature range apparently available, Macfarlane and Angell [3] carried out the deposition according to Olander and Rice's prescription directly into a DSC pan, so that precise quantitative measurements of the heat capacity jump could be made. However, after an initial positive result was traced to vacuum pump oil incorporated in the deposit, they were unable to detect any thermal effect at all before crystallization. The sensitivity of the measurement was much greater than needed to detect either the jump reported in ref. 10, or the smaller jump anticipated from solution data extrapolations [12]. Accordingly, MacFarlane and Angell concluded that either the glass transition did not exist (analog of the "tetrahedral" glassformer, $BeF_2$ [13]), or that it lay above the crystallization threshold - as for many hyperquenched metallic glasses.

Then Bruegeller and Mayer [14] and Dubochet and McDowell[15] almost simultaneously reported the preparation of glassy water by rapid quenching of microscopic samples of water into a cryogenic liquid . Bruegeller and Mayer identified the formation of a glassy state by the subsequent release of the heat of crystallization of the quenched substance, though again no actual glass transition was detected. Dubochet and McDowell detected it by the diffuse diffraction rings in electron microscope studies. Mayer and coworkers [16] subsequently determined that the glassy phase obtained in this manner was likely to be contaminated by inclusion of molecules of the cryoquench liquid in the vitreous water network. To avoid this Mayer [17] developed the aerosol droplet hyperquench method in which hypersonic droplets formed by vacuum expansion, were splatted against a liquid nitrogen-cooled substrate. This conglomerate of tiny quenched droplets could then be studied microscopically, spectroscopically [4b] or calorimetrically [9,18]. as desired. This form of glassy water is as pure as it is possible to obtain. Subsequent studies of this conglomerate [18] showed that, after sufficient annealing at temperatures near 120K, it yielded a weak and spread-out DSC endotherm commencing at about 136K. It was assigned to the glass transition temperature for vitreous water A similar behavior was then observed, and interpreted as a glass transition, for ASW [19].

The calorimetric results from the detailed report of Hallbrucker and Mayer on the hyperquenched glass [9] are reproduced in Fig. 1. An important feature of the data is the release of heat starting at 120K and continuing until the crystallization commences 155K. This exotherm was interpreted by Hallbrucker and Meyer, correctly we believe, in terms of the relaxation of the high energy quenched state towards lower enthalpy states characteristic of slowly cooled glasses. Such annealing exotherms are well known in the calorimetry of glasses cooled at rates in excess of the reheating rate [20]. As would be expected, the effect is larger, both in magnitude and temperature range, the larger the difference in heating and cooling rates. Indeed, this is predicted by the Adam-Gibbs equation[21] applied to non-equilibrium systems [22] .We now briefly explain the usefulness of such exotherms in the study of glasses before returning to the significance of the Fig. 1 result.

When the cooling rates are known, the excess heats over that of a "standard scan," i.e. one obtained at equal cooling and heating rates, can be used to determine the fictive temperature (the temperature at which equilibrium was lost during cooling) as a function of cooling rate. The results can be used to determine the activation energy for the enthalpy relaxation process [20,22].

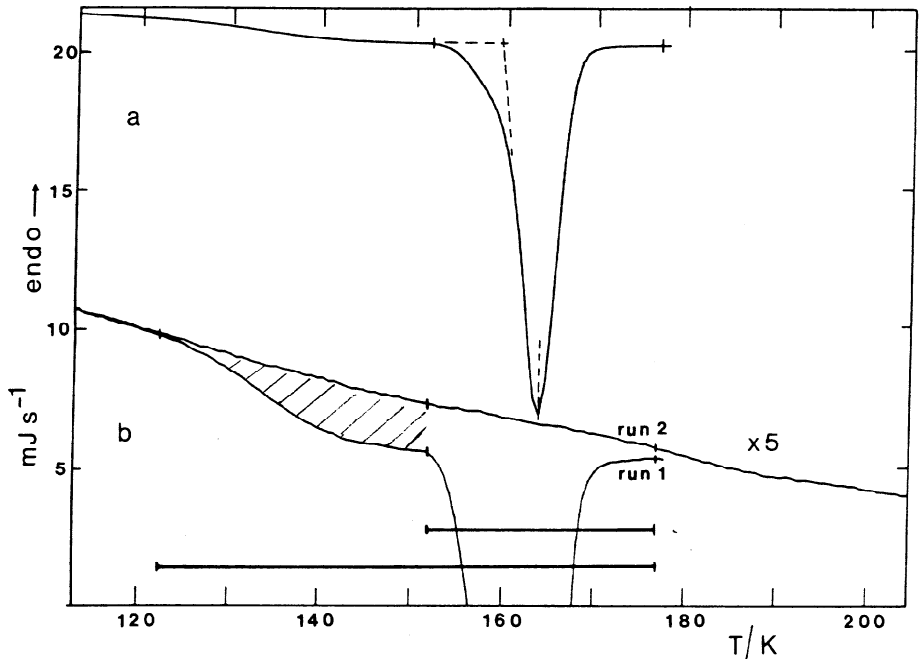

Fig. 1. Initial, and second, differential scanning calorimeter upscans of hyperquenched glassy water, showing (hatched area) the exothermic release of enthalpy stored in the hyperquenched state. (adapted from ref. 20, by permission)

Related measurements have subsequently been used by Mayer and co-workers to study water in hydrogels [23]. Here we show how it applies to the model fragile liquid glassformer, o-terphenyl OTP, and how it can be used to assess the cooling rates of hyperquenched glassy OTP [24]. Fig. 2 shows the scans for different cooling and fixed heating rates, and Fig. 3 shows the plot of log $Q/Q_0$ vs scaled fictive temperature $T_F^0/T_F$, where $T_F^0$ is the fictive temperature at the standard scan rate $Q^0$. The slope of this plot is the so-called "m" fragility index [25] or "steepness index" [26]. It is related to the activation energy for enthalpy relaxation by

$$m = \Delta H/2.303RT_g \qquad (1)$$

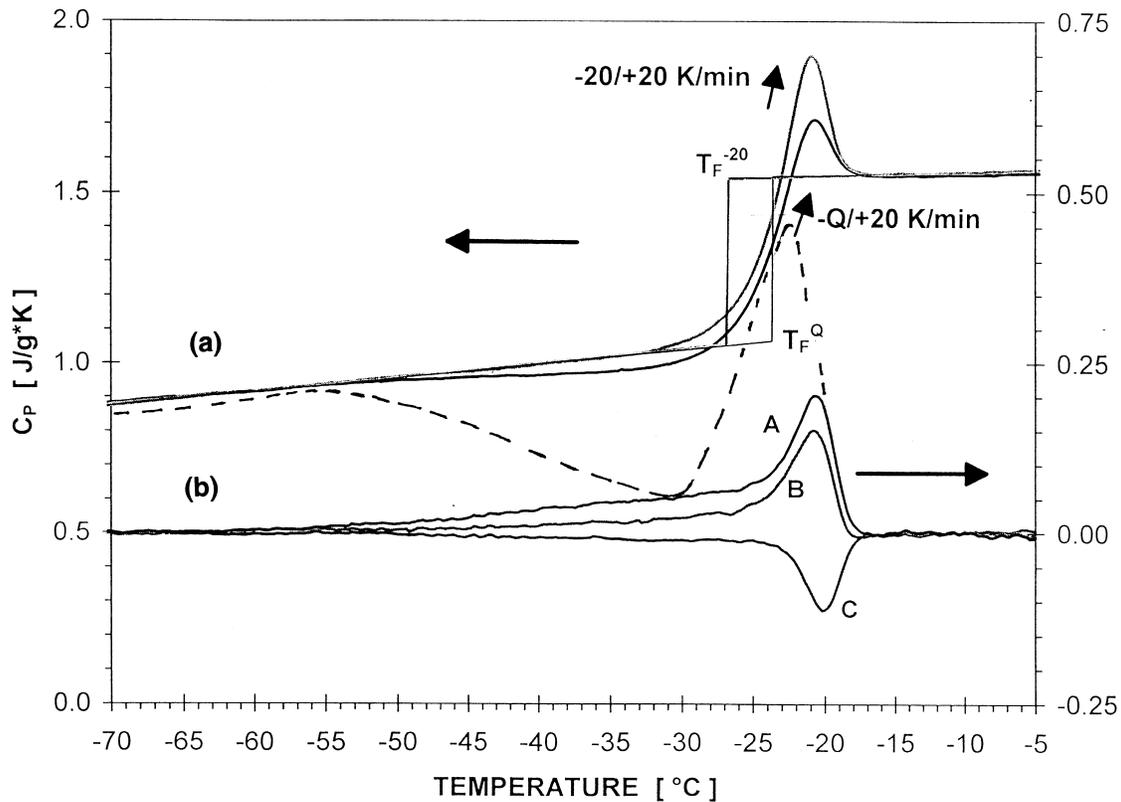

Fig. 2. (a) DSC upscans at the ("standard") heating rate, 20K/min, of two o-terphenyl OTP glasses formed at different cooling rates, –20K/min, and a -Q K/min, which is the maximum rate permitted by the instrument. The upper scan is called the "standard scan".

(b) (right hand ordinate) The difference between the two curves of part (a) is shown as curve A. This may be called the "excess heat capacity" curve, and appears as a positive quantity. Its integral gives the excess enthalpy of the quenched glass. Curve B is the difference between the standard scan and the 20K/min upscan following the fastest "controlled temperature " downscan permitted by the DSC instrument, which is –73K/min. Curve C is obtained by differencing a cooling scan at –10K/min, less than the standard value, with the standard scan, and of course it shows a negative displacement. The difference areas (excess enthalpies) are used to obtain the fictive temperatures that are plotted in Fig. 3. The dashed curve is the excess heat capacity obtained for the hyperquenched OTP sample, which was quenched at a rate some four orders of magnitude faster than any of the others, hence has a much greater excess enthalpy .

The m fragility obtained for OTP by the slope of Fig. 3 is 77, in excellent agreement with the value of 76 obtained from viscosity data [27] and with the value 81 obtained from dielectric relaxation data [25(b)].

The Fig. 3 plot can serve as a basis for calculating the cooling rates of hyperquenched glasses. Since the viscosity of OTP follows an Arrhenius law over several orders of magnitude near $T_g$ [27], it is reasonable to use the Fig. 3 Arrhenius plot, in combination with the fictive temperature determined calorimetrically for a rapidly quenched glass, to determine the quenching rate by which that glass was formed. For instance, we find

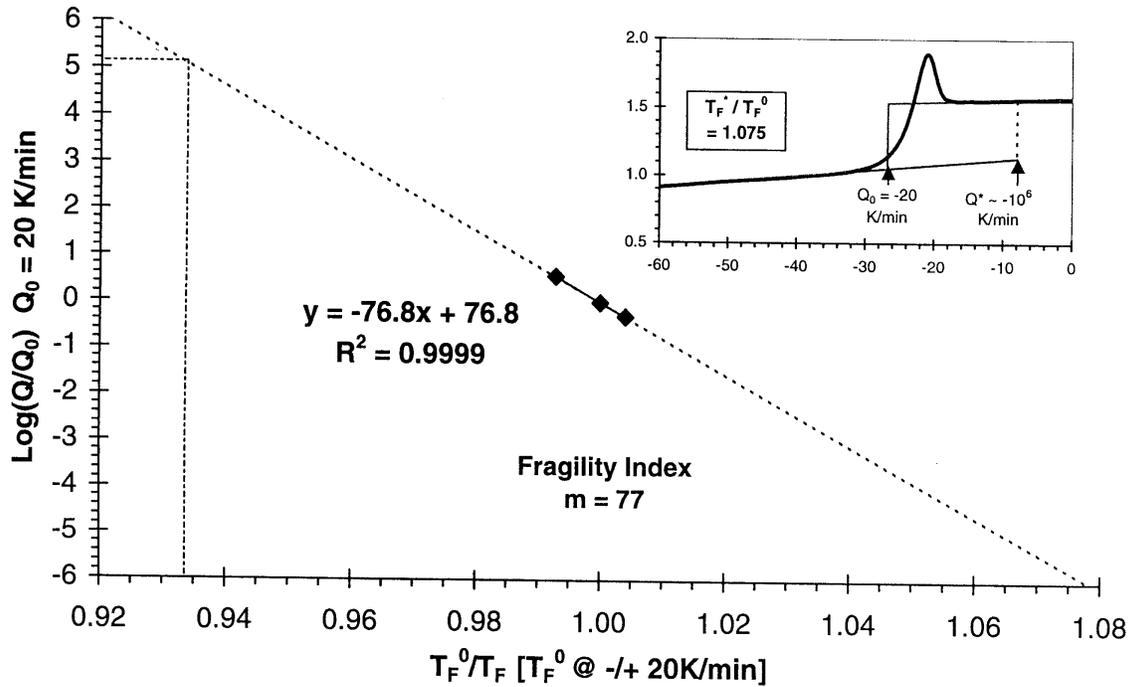

Fig. 3. Scaled Arrhenius plot of the cooling rate for glasses formed at -Q K/min vs resulting fictive temperature. The scaling parameters are the standard cooling rate ($Q_0$ = -20K/min) and fictive temperature $T_F^0$ of the standard glass, formed by $Q_0$ cooling. The slope of this plot gives the m fragility index directly. The obtained value 77 is to be compared with the value 76 obtained from the activation energy for viscosity at $T_g$ [26] using Eq. (1), and the value 81 obtained from dielectric relaxation data [24(b)]. The vertical dashed line is drawn at the scaled fictive temperature of hyperquenched OTP, which is assessed in the insert (see below) The quench rate for this sample can therefore be obtained as $Q = 10^5 Q_0$ which is $2 \times 10^6$ K/min, or $6.6 \times 10^4$ K/sec.

**Insert** shows the definition of the fictive temperature [22,23] for the standard scan, and the fictive temperature for the hyperquenched OTP glass. The latter is assessed from the upscan exotherm enthalpy, which is obtained from the excess heat capacity curve displayed in Figs 2 and 4. As seen in the insert, the fictive temperature of the standard scan coincides (within 0.2K) with the glass transition temperature defined by the "$C_p$ onset" criterion [23] that the quenching rate for the sample used to obtain Fig. 3 points, but quenched at the maximum rate possible in the DSC, has a cooling rate of 247 K/min, as would be expected from the manufacturer's description. The dashed line in Fig. 3 is placed at the fictive temperature of an OTP sample which was quenched into a liquid nitrogen-cooled DSC pan, using an electrospraying technique [24,28], and subsequently upscanned at 20K/min to obtain the equivalent, for OTP, of Fig. 1 for water. The Fig. 3 plot shows that the effective cooling rate for this sample was $2 \times 10^6$ K/min, about the same as that attributed to water hyperquenched by the aerosol splat technique [17]. The much larger upscan exotherm for this hyperquenched glass is indicated by the dashed curve in Fig. 2, and it is noted now that the exotherm has a maximum well below $T_g$.

Examination of upscan exotherms from cases of roller-quenched or melt-spun metallic glasses, show similar exotherms with maximum excess heat release rates [24,29]. This is also found in a our study [24] of a melt-spun pitch (used as graphite fiber precursor ) in which the cooling rate is estimated to be $10^7$K/min [30]. In Fig. 4 we collect all these data into a master-plot using a $T_g$-scaled temperature axis, and include the data from Fig. 1 as a dotted line by using the generally accepted value for the glass transition temperature, 136K [14,18,19]. It is immediately clear that the hyperquenched water plot is out of place, relative to the others.

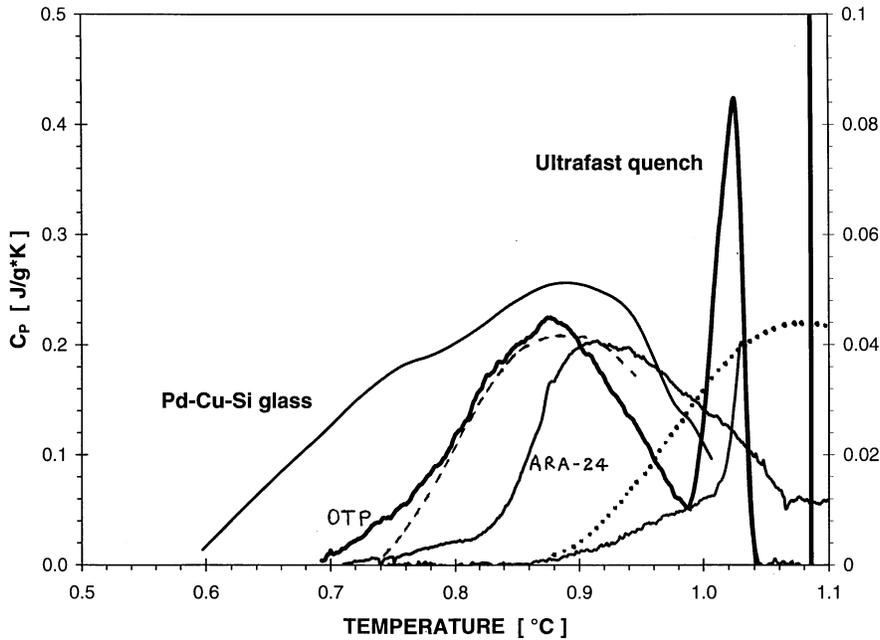

Fig. 4. Comparison of the excess heat capacities (excess over the standard scan) vs Tg-scaled temperature, of three hyperquenched bulk glass-formers [23], with that from the study of hyperquenched water (Fig. 1, ref. 9). The latter are plotted twice, firstly (dotted line) for the case in which the scaling temperature is taken as the accepted $T_g$ value 136K [31,34], and secondly (dashed line) for the choice $T_g$ = 165K. The bulk glassformers, for which $T_g$ can be measured directly, are (i) a $Pd_{77.5}Si_{16.5}Cu_6$ glass which was melt-spun at an estimated cooling rate of $10^6$ K/s [29], OTP electrospray-quenched at $7\times10^4$K/s according to Fig. 3, and a Mitsubishi ARA-24 pitch [24], melt-spun at an estimated $10^5$K/s quench rate.

There are two possible reasons for this. The first is that the excess enthalpy relaxation process for water is not like that of a fragile liquid as exemplified by OTP and the metallic glass $Pd_{77.5}Si_{16.5}Cu_6$ [29, 32]. This would conflict with recent claims, based on difficult diffusivity measurements near the crystallization temperature [33], that water is a very fragile liquid (instead of a strong liquid in this temperature range, as claimed in refs. 34).

The second possibility, which is more strongly indicated and which contradicts both refs. 33 and 34, is that water in the range 130-155K is not a liquid of any sort. Rather it is a non-diffusive glass, as argued by several authors on the basis of unrelated experiments [14, 35,36]. The glass transition for water, according to these authors must fall well above the crystallization temperature, and therefore cannot be observed directly. However, fortunately the information on where it might lie is stored in the quenched-in enthalpy whose release is shown in Fig.1.

To estimate where this unobservable (on long time scale measurements) $T_g$ might lie, we look for a scaling temperature that superposes the hyperquenched water upscan exotherm data onto the other systems' exotherms. The dashed curve in Fig. 4 that matches the data for OTP quite well, is obtained when the excess enthalpy of Fig. 1 is temperature-scaled using the value $T_g = 165K$. Interestingly enough, this is the temperature of the glass transition observed for water in nanodroplet inclusions in the hydrogels studied by Mayer and co-workers [23,37]. It is also the temperature predicted recently [38] by an Adam-Gibbs equation fit to the high temperature viscosity data on supercooled water which was extended to low temperatures using a thermodynamically constrained estimate of the excess entropy variation between 150 and 235K.

What, then, is the origin of the weak thermal effects observed, and reported as glass transitions, by Johari et al [18,19], and apparently confirmed by Klug and Handa [39]? As suggested by Fisher and Devlin [35], these may be the relaxation of the non-diffusive Bjerrum defects that are also seen in Ice $I_h$. at about the same temperature, and were accelerated by impurity additions in the work of Suga and co-workers [40] to the point where most of the residual entropy of ice could be lost in a first order phase transition. These are currently being characterized in great detail by NMR relaxation methods [B. Geil and F. Fujara, private communication] which show them also to be decoupled from diffusion. Such modes might be thought of as the equivalent of the Johari-Goldstein relaxations familiar in other molecular glass-formers [41], which would explain why they exhibit simple Arrhenius behavior [34(b)]. Is this sort of decoupled reorientation mechanism also responsible for the dielectric relaxation seen in nanodroplet water imbibed in poly-(2-hydroxyethyl-methacrylate (polyHEMA) near its glass transition at 162K (the value given for a 34 wt% water sample [37])? These data are also Arrhenius in character with an m fragility of only 25 [42], and are faster than the process determining the glass transition at 162K. The latter has a higher (but still relatively low [43]) m fragility of 39 (activation energy of 120kJ/mole[37]). These are questions which remain to be answered. The light is visible at the end of the tunnel, but remains a little dim.

**Acknowledgements:**
 This work was carried out under the auspices of the NSF, Solid State Chemistry Program, Grant Nos. DMR -9614531, and DMR-0082535

**Figure captions**

Fig. 1.
 Initial, and second, differential scanning calorimeter upscans of hyperquenched glassy water, showing the exothermic release of enthalpy stored in the hyperquenched state. (from ref. 20, by permission)

Fig. 2.
(a) DSC upscans at the ("standard") heating rate, 20K/min, of two o-terphenyl OTP glasses formed at different cooling rates, –20K/min, and a -Q K/min, which is the maximum rate permitted by the instrument. The upper scan is called the "standard scan".

(b) (right hand ordinate) The difference between the two curves of part (a) is shown as curve A. This may be called the "excess heat capacity" curve. Curve B is the difference between (excess heat capacity of) the standard scan and the 20K/min upscan following the fastest "controlled temperature " downscan permitted by the DSC instrument, which is –73K/min. Curve C is obtained by differencing a cooling scan at –10K/min, less than the standard value, with the standard scan, and of course shows a negative displacement. The difference areas are used to obtain the fictive temperatures that are plotted in Fig. 3.

Fig. 3
Scaled Arrhenius plot of the cooling rate for glasses formed at-Q K/min vs resulting fictive temperature. The scaling parameters are the standard cooling rate ($Q_0$ = -20K/min) and fictive temperature $T_F^0$ of the standard glass, formed by $Q_0$ cooling. The slope of this plot gives the m fragility index directly. The obtained value 77 is to be compared with the value 76 obtained from the activation energy for viscosity at Tg [26]using Eq. (1), and the value 81 obtained from dielectric relaxation data [24(b)]. The vertical dashed line is drawn at the scaled fictive temperature of hyperquenched OTP, which is assessed in the insert (see below) The quench rate for this sample can therefore be obtained as $Q = 10^5 Q_0$ which is $2 \times 10^6$ K/min, or $6.6 \times 10^4$ K/sec.

**Insert** shows the definition of the fictive temperature [22,23] for the standard scan, and the fictive temperature for the hyperquenched OTP glass. The latter is assessed from the upscan exotherm enthalpy, which is obtained from the excess heat capacity curve displayed in Fig. 4. As seen in the insert, the fictive temperature of the standard scan coincides (within 0.2K) with the glass transition temperature defined by the "$C_p$ onset" criterion [23]

Fig. 4.
Comparison of the excess heat capacities (excess over the standard scan) of three hyperquenched bulk glass-formers [23], with that from the study of hyperquenched water (Fig. 1, ref. 20). The latter are plotted twice, firstly (dotted line) for the case in which the scaling temperature is taken as the "accepted" value 136K [31], and secondly (dashed line) for the choice $T_g = 165K$. The bulk glassformers, for which $T_g$ can be measured directly, are (i) Pd-Cu-Si which was melt-spun at an estimated cooling rate of $10^6$ K/s [29], OTP electrospray-quenched at $7 \times 10^4$ K/s according to Fig. 3, and Mitsubishi ARA-24 pitch melt-spun at an estimated $10^5$ K/s quench rate.